\documentclass[12pt]{article}
\usepackage{orcidlink}
\usepackage{float}
\usepackage{amsmath,amssymb}
\usepackage{comment}
\usepackage{bm}
\usepackage{color}
\usepackage{cite}
\usepackage{graphicx}
\usepackage{epstopdf}
\usepackage[dvipdfmx]{epsfig}
\usepackage{epsfig}
\usepackage{braket}
\usepackage{float}
\usepackage{color}
\usepackage{hyperref}
\newcommand{\tmop}[1]{\ensuremath{\operatorname{#1}}}

\newcommand{\bea}{\begin{eqnarray}} 
\newcommand{\eea}{\end{eqnarray}}

\begin{document}
\title{
\begin{flushright}
\ \\*[-80pt] 
\begin{minipage}{0.2\linewidth}
\normalsize
%arXiv:YYMM.NNNN \\
%XXXX-0000 \\*[50pt]
\end{minipage}
\end{flushright}
{\Large \bf 
Thermal Casimir Effect in A Schwarzschild-like Wormhole Spacetime 
\\*[20pt]}}

\author{ 
\centerline{
\begin{minipage}{\linewidth}
\begin{center}
Arista Romadani\orcidlink{0000-0002-2746-7204}$^{1,2}$\footnote{\texttt{E-mail address: arista.romadani@uin-malang.ac.id}}, 
Apriadi Salim Adam\orcidlink{0000-0001-6587-5156}$^{3,4}$\footnote{\texttt{E-mail address: apriadi.salim.adam@brin.go.id}}, 
Ar Rohim\orcidlink{0000-0002-6629-9979}$^{4}$\footnote{\texttt{E-mail address: ar.rohim@ui.ac.id}}, \\
Bintoro Anang Subagyo\orcidlink{0000-0001-8350-6725}$^1$\footnote{\texttt{E-mail address: bintoro.subagyo@its.ac.id}}, and
Agus Purwanto\orcidlink{0000-0002-4103-3613}$^1$\footnote{\texttt{E-mail address: purwanto@physics.its.ac.id}}
\end{center}
\end{minipage}}
\\*[20pt]
\centerline{
\begin{minipage}{\linewidth}
\begin{center}
$^1${\it \normalsize 
Department of Physics, FSAD, Sepuluh Nopember Institute of Technology, Kampus ITS Sukolilo, Surabaya, 60111, Indonesia} \\*[5pt]
$^2${\it \normalsize
Department of Physics, Faculty of Science and Technology, Universitas Islam Negeri Maulana Malik Ibrahim Malang, Malang  65144, Indonesia} \\*[5pt]
$^3${\it \normalsize
Asia Pacific Center for Theoretical Physics, Pohang 37673, South Korea} \\*[5pt]
$^4${\it \normalsize
Research Center for Quantum Physics, National Research and Innovation Agency (BRIN)\\
South Tangerang 15314, Indonesia} \\*[5pt]
$^4${\it \normalsize
Departemen Fisika, FMIPA, Universitas Indonesia, Depok 16424, Indonesia} \\*[5pt]
\end{center}
\end{minipage}}
\\*[50pt]}

\date{} % Kosongkan bagian tanggal

\begin{titlepage}
\maketitle
\thispagestyle{empty}

\begin{center}
{\bf \small Abstract}
\end{center}

\begin{minipage}{0.9\linewidth}
\small
We study the finite‑temperature Casimir effect for a massless scalar field confined between two parallel plates in a Schwarzschild‑like wormhole spacetime. Imposing Dirichlet boundary conditions, we compute the renormalized Casimir free energy in the comoving frame. We find that the thermal correction to the renormalized Casimir free energy decreases gradually with the temperature and becomes independent of the background geometry in this frame. Thermodynamic quantities derived from the Casimir free energy, namely, the renormalized Casimir entropy, internal energy, and heat capacity at constant volume, exhibit distinct temperature dependence. At low temperatures, all thermodynamic quantities recover the expected behavior, consistent with the fundamental laws of thermodynamics. These results provide a compact framework for analyzing quantum vacuum forces in gravitational backgrounds.
\end{minipage}
\end{titlepage}

%%%%%%%%%%%%%%%%%%%%%%%%%%%%%%%%%%%%%%%%%%%%%%%%%%%%%%%%%%%%
%%%%%%%%%%%%%%%%%%%%%%%%   Introduction   %%%%%%%%%%%%%%%%%%%%%%%%%%%
%%%%%%%%%%%%%%%%%%%%%%%%%%%%%%%%%%%%%%%%%%%%%%%%%%%%%%%%%%%%
\section{Introduction}
\label{sec1}
The Casimir effect arises from the modification of quantum vacuum fluctuations by boundary conditions, geometry, topology, or background fields. In its original form, it describes an attractive force between two perfectly parallel plates due to fluctuations in the electromagnetic vacuum \cite{Casimir1948}. From van der Waals theory, Casimir extended the dispersion-force calculations between two parallel ideal conducting plates, leading to the Casimir effect phenomenon \cite{Klimchitskaya:2015}. Experimental investigations have consistently validated the Casimir effect, showing strong agreement with theoretical predictions \cite{Sparnaay1958}. Recent research has provided high-precision evidence supporting its existence \cite{Lamoreaux1997, Mohideen:1998, Roy:1999, Bressi:2002}. In addition, the Casimir effect has various applications in nanotechnology \cite{Bellucci:2009} and condensed matter physics \cite{Grushin:2011a}. 

The Casimir effect can occur not only in the electromagnetic field but also in scalar \cite{Cruz:2020, Fendley:1994, Bordag:1997}, vector \cite{teo:2010}, and fermion fields \cite{Chodos:1974a,Chodos:1974b, Johnson:1975zp}. The behavior of each field depends on the plate geometry specified by the boundary conditions. For a scalar field, Dirichlet boundary conditions (field vanishing at the plate's surface) can be applied, and Neumann or mixed boundary conditions are also viable \cite{Ambjorn:1983, sasanpour:2023, Cruz:2017, Erdas:2025}. However, fermion fields cannot fulfill such boundary conditions because their solutions come from a first‑order differential equation; instead, one employs a bag boundary condition, commonly known as the MIT bags, that pushes the flux to vanish at the plate's surface \cite{Chodos:1974a,Chodos:1974b,Johnson:1975zp}. Extensions that incorporate a chiral angle have also been studied \cite{Jaffe:1989pn,Theberge:1980ye,Lutken:1983hm,Chodos:1975ix,Oikonomou:2009zr,Zahed:1984jq,Rohim:2021saa,Nicolaevici:2016vnj,Chernodub:2017ref,Chernodub:2016kxh,Rohim:2022mri}.

Several attempts have also been made to study the influence of the background geometry on the Casimir effect, as curvature and topology could alter the spectrum of quantum vacuum modes and therefore the renormalized vacuum energy \cite{Bezerra:2014a, Bezerra:2016}. These include gravitational backgrounds such as Kerr \cite{sorge:2014}, Schwarzschild \cite{sorge:2022, sorge:2019a}, and wormhole spacetimes \cite{Krasnikov:2018, Sorge:2020} (see also Refs. \cite{sorge:2005, napolitano:2008, sorge:2009, sorge:2019, lima:2019}). This connection is particularly relevant in wormhole physics, where Casimir-induced negative energy can help maintain traversable wormhole geometries from collapsing \cite{Khabibullin:2006, Visser:1989}. In classical general relativity, static traversable wormholes typically require a violation of the null energy condition near the throat, a condition typically associated with an exotic matter. In this context, Casimir energy, which can become negative in appropriate configurations, has been widely regarded as an adequate quantum source of the exotic stress energy needed to support such geometries.

Another attractive type of wormhole is the Schwarzschild-like wormhole. It provides a slight generalization of the traversable Schwarzschild wormhole spacetime by introducing a specific linear shape function. This type of wormhole is also studied as "a super accelerator", since its matter source is confined to a compact neighborhood of the throat, and the $tt$-component (in the Schwarzschild coordinates) of its metric has a single minimum \cite{Krasnikov:2018nga}. These two properties could make it a good candidate as an accelerator. On top of that, the Schwarzschild-like wormhole has also been explored in the context of asymptotically safe gravity \cite{Alencar:2021}.

The impact of temperature on Casimir energy/force is understudied, particularly for the context of wormhole physics. Although Casimir energies at zero temperature have been studied in wormhole geometries \cite{Santos:2021, sorge:2014, Sorge:2020}, the finite-temperature scalar field confined in an orbiting cavity around a Schwarzschild-like wormhole has not yet been systematically explored. This question is particularly compelling given that thermal fluctuations can significantly alter and in some regimes dominate zero-temperature quantum vacuum contributions in curved spacetime \cite{zhang:2015, munizhotcas:2025, Sushkov:2011, Bostrom:2000}. Furthermore, temperature effects may play a crucial role in traversable wormhole physics, especially in models involving hot Casimir wormholes \cite{hotcas:2025}. In particular, in their model, it has been shown that the thermal effect on the Casimir energy could modify the size of the wormhole throat. However, this investigation lies beyond the scope of our study.

Motivated by the above considerations, in this paper, we study the thermal Casimir energy in a wormhole spacetime, focusing on a massless scalar field confined between two parallel plates orbiting a static, zero‑tidal Schwarzschild‑like wormhole. Imposing Dirichlet boundary conditions, we compute the vacuum energy in the comoving frame. It turns out that the resulting vacuum energy and its correction are divergent, and one should perform  appropriate regularization and renormalization to obtain the renormalized Casimir energy. Then, we examine how the temperature and background geometry influence this renormalized Casimir energy. We also compute the corresponding thermodynamic quantities—entropy, internal energy, and heat capacity at constant volume—from the renormalized Casimir energy. 

The rest structure of this paper is organized as follows. In section \ref{sec2}, we describe the model of our system, namely, a massless scalar field confined between two parallel plates orbiting a Schwarzschild-like wormhole. We then define the local reference frame of the orbiting observer comoving with the Casimir apparatus. In section \ref{sec3}, we solve the scalar field equation and apply the Dirichlet boundary conditions to calculate the Casimir energy density around the Schwarzschild-like wormhole. In section \ref{sec4}, we investigate  the finite-temperature of the Casimir energy density and calculate thermodynamic quantities such as entropy, internal energy, and heat capacity at a constant volume. Finally, section \ref{sec5} is devoted to our summary. In this paper, we will use a Planck unit, $c=\hbar=k_B=G=1$.
%%%%%%%%%%%%%%%%%%%%%%%%%%%%%%%%%%%%%%%%%%%%%%%%%%%%%%%%%%%%%%%%%%%%%%%%%%%%%%%%%%%%

\section{The Model}
\label{sec2}
\begin{figure}[!ht]
    \centering
    \includegraphics[width=0.4\linewidth]{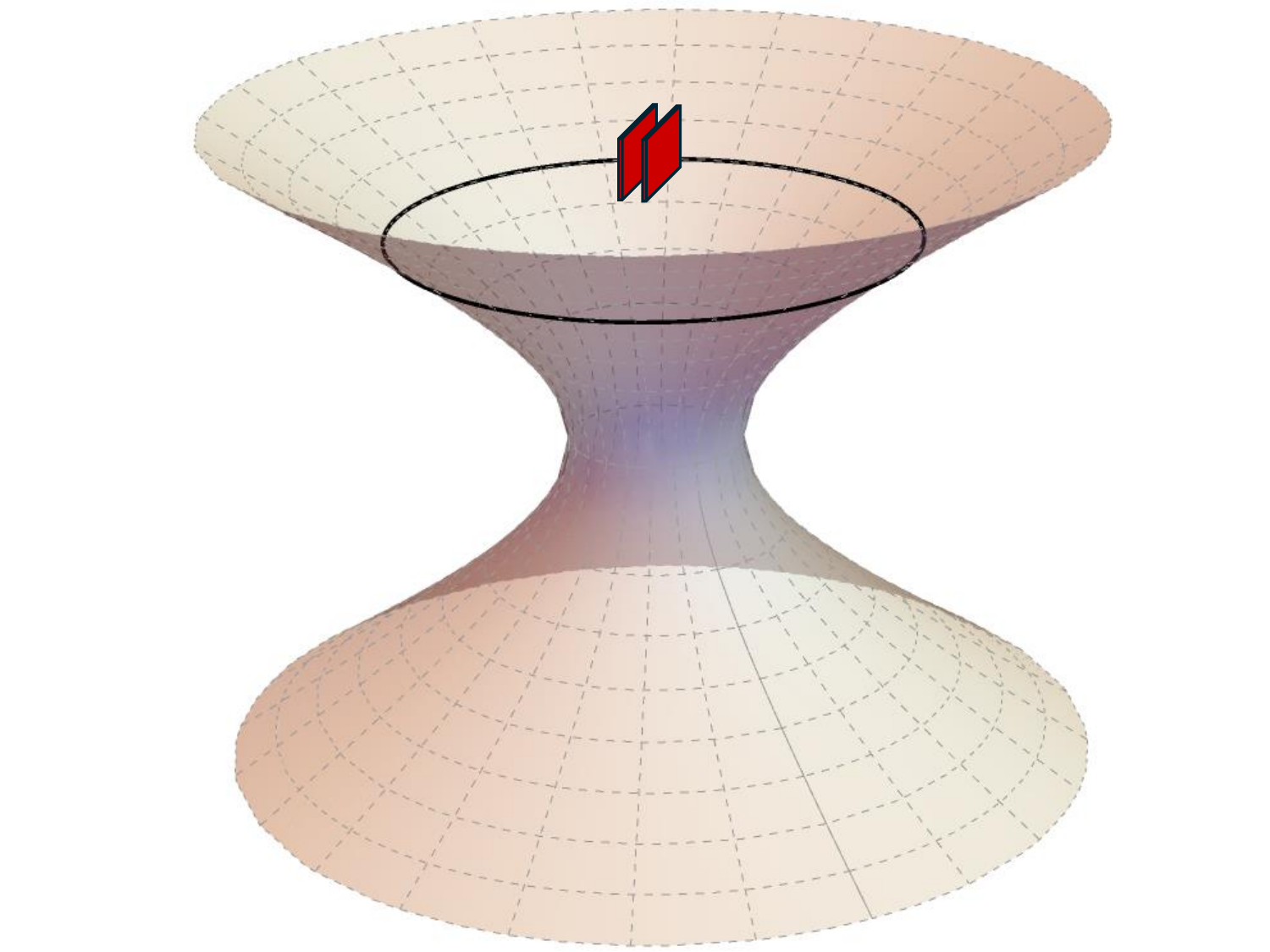}
    \caption{Casimir cavity orbiting around Schwarzschild-like wormhole}
    \label{fig0}
\end{figure}

The present study considers the massless scalar field confined between two parallel plates placed at $z=0$ and $z=L$ in the vacuum cavity. The normal surface of the plates is parallel to the $z$-axis. The Casimir cavity moves around a static and zero tidal Schwarzschild-like wormhole in a circular orbit, as shown in Figure \ref{fig0}. We assume that this Casimir cavity is an exotic matter in spacetime that could generate a negative energy to prevent the throat from collapsing. Tidal effects inside the cavity are negligible, and only the gravitational effects are induced by the wormhole geometry. Introducing $S$ as the area of each plate, the following inequalities hold:
\begin{eqnarray}
    L \ll \sqrt{S} \ll r_0 \leq r, \label{model}
\end{eqnarray}
where $r_0$ is the throat radius and the radial coordinate satisfies the constraint $r > r_0$. Therefore, the length of the plates is much less than their radial coordinate, $L\ll r$. %and we do not consider tidal effects inside the cavity. 
The geometry of the plates is described by the Dirichlet boundary conditions, which require the vanishing field $\phi(t,x,y,z)$ on the plate's surface as follows \cite{Ambjorn:1983}:
\begin{eqnarray}
    \phi(t,x,y,0)=\phi(t,x,y,L)=0, \label{dirichlet}
\end{eqnarray}
where the cavity is in a locally comoving referential frame with Cartesian coordinates defined on one of the plates and the $z$-axis is tangential to the circular orbit path.

\subsection{Schwarzschild-like Wormhole Spacetime}
A static spherically symmetric wormhole is described by Morris-Thorne $(t,r,\theta,\phi)$ with the metric \cite{Morris:1988}:
\begin{eqnarray}
    ds^2=e^{2\Phi(r)} dt^2-\frac{dr^2}{1-\frac{b(r)}{r}}-r^2d\theta^2-r^2\sin^2 \theta d\phi^2, \label{a1}
\end{eqnarray}
where $e^{\Phi(r)}$ and $b(r)$ are the redshift and shape functions, respectively. Meanwhile, the angles $\theta$ and $\phi$ are defined within the ranges $0 \le \theta \le \pi $ and
$0 \le \phi\le 2\pi$. The Schwarzschild black hole solution allows the formation of a Schwarzschild wormhole. However, this wormhole is not traversable due to the event horizon. By removing the event horizon, the wormhole becomes stable and the Schwarzschild wormhole can be traversable. This can only be done by taking the redshift function of its radial metric component as a linear function \cite{Cataldo:2017}. The method produces a metric known as a Schwarzschild-like wormhole, which is a static Schwarzschild black hole with no horizon or traversable \cite{Santos:2021}.

This paper considers a zero-tidal spacetime where the redshift function is equal to unity, that is, $\Phi(r)=0$. The shape function is linearly dependent on the radial coordinate, $b(r) = \alpha+\gamma r$, with $\alpha$ and $\gamma$ being arbitrary constants. By applying the condition of Schwarzschild, $b(r_0)=r_0$, we obtain that the wormhole shape function has a minimum at $r = r_0$, where the wormhole throat is located. In addition, we find that $\alpha=r_0(1-\gamma)$ and the shape function takes the form $b(r)=(1-\gamma)r_0+\gamma r$. As a results, in Eq.\eqref{a1}, the radial coordinate covers the range $r_0<r<\infty$. This condition leads to a metric so-called the Schwarzschild-like wormholes and our metric now becomes \cite{Alencar:2021}:
\begin{eqnarray}
    ds^2= dt^2-\frac{dr^2}{\left(1-\gamma\right) \left(1-\frac{r_0}{r}\right)}-r^2d\theta^2-r^2\sin^2 \theta d\phi^2. \label{a2}
\end{eqnarray}
The metric Eq. \eqref{a2} requires $0<\gamma < 1$ to ensure that the radial component of the metric will not diverge and describes non-asymptotically flat wormholes. In the particular case $\gamma=0$, since $b(r)/r \to 0$ as $r\to \infty$, the metric becomes asymptotically flat. Therefore, the spacetime approaches the Minkowski form at large distances and reduces to the standard Morris–Thorne wormhole geometry \cite{Cataldo:2017}.

\subsection{Isotropic Coordinate}
In this part, we will transform Eq.\eqref{a2} into isotropic coordinates $(t,\rho,\theta,\phi)$, following the procedure in Ref.\cite{Sorge:2020}. The formulation in isotropic coordinates is essential for obtaining the metric in the Minkowskian or comoving frame. Using this frame allows us to calculate the vacuum energy (or Casimir energy) by adjusting the geometry of parallel conductor plates rotating in a Schwarzschild-like wormhole. 

Typically, the isotropic metric is expressed as follows
\begin{eqnarray}
d s^2=d t^2 - F^2(\rho) [d \rho^2 + \rho^2 d \theta^2 + \rho^2 \sin^2 \theta d \phi^2], \label{b0}
\end{eqnarray}
where $F(\rho)$ is a conformal factor to be determined later. By comparing the spatial line element of Eqs.\eqref{a2} and \eqref{b0}, we have
\begin{align}
    r&=\rho F(\rho), \label{radial1} \\
    \frac{dr}{d\rho}&=F(\rho) \sqrt{\left(1-\gamma \right) \left( 1 - \frac{r_0 }{r}\right)}\label{radial2}.
\end{align}
By solving Eqs.\eqref{radial1} and \eqref{radial2}, we obtain the conformal factor $F(\rho)$ as follows\footnote{We may compare the expression of $F(r)$ in this work and that of $A(r)$ in Ref.\cite{Sorge:2020}. In this case, $F(r)$ will be reduced to $A(r)$ if one sets $r_0=-4\rho(1\mp \sqrt{A(r)})$.}:
%By solving Eqs.\eqref{radial1}, \eqref{radial2}, and setting $\gamma = 0$, $r_0 = 2M$ to determine the integration constant, we obtain the conformal factor $F(\rho)$ as follows:
\begin{eqnarray}
  F(\rho) = \frac{\left( \frac{\rho}{r_0} \right)^{\sqrt{(1 - \gamma) } - 1}}{1
  - \gamma} \left( 1 + \frac{1 - \gamma}{4 \left( \frac{\rho}{r_0}
  \right)^{\sqrt{(1 - \gamma) }}} \right)^2, \label{a3}
\end{eqnarray}
where the integration constant is determined by requiring that the metric reduces to the isotropic Schwarzschild solution for $\gamma = 0$ and $r_0 = 2M$. %\footnote{\textcolor{red}{The relation between $F(r)$ in this work and $A(r)$ in Ref.\cite{Sorge:2020} satisfy $r_0=-4\rho(1\mp \sqrt{A(r)})$.}}
Substituting Eq.\eqref{a3} into \eqref{b0} leads to the following metric:
\begin{eqnarray}
  d s^2= d t^2 - \frac{\left( \frac{\rho}{r_0} \right)^{2\left({\sqrt{(1 - \gamma) } - 1}\right)}}{(1
  - \gamma)^2} \left( 1 + \frac{1 - \gamma}{4 \left( \frac{\rho}{r_0}
  \right)^{\sqrt{(1 - \gamma) }}} \right)^4 [d \rho^2 + \rho^2 (d \theta^2 + \sin^2 \theta d \phi^2)]. \label{a4}
\end{eqnarray}
Note that the above metric is the suitable form when we introduce the Casimir apparatus's parallel plate geometry using Cartesian coordinates. It can be realized by changing from isotropic Schwarzschild-like wormhole to the comoving frame.

The throat of the metric in Eq.\eqref{a4} is defined by the minimum of the areal radius $R=\sqrt{-g_{\theta \theta}}$. It can be obtained by setting $dR/d\rho=0$ and gives the throat at $\rho_{\text{thr}}=r_0\left(\frac{1-\gamma}{4}\right)^{1/\sqrt{1-\gamma}}$. This minimum value implies that $R(\rho_{\text{thr}})=r_0$. In addition, we also analyze the singularity from the Kretschmann scalar $K=R_{\mu\nu\alpha\beta}R^{\mu\nu\alpha\beta}$ at $\rho=0$, $\rho \to \infty$, and $\rho=\rho_{\text{thr}}$. As a result, there is no curvature singularity for $\rho>0$ because the Kretschmann scalar is finite. The true singularity happens in the limit $\gamma \to 1$ where the metric is divergent for an arbitrary value of $\rho$.

\subsection{The Circular Orbits}
In this section, we study the circular orbits of the Casimir cavity around the Schwarzschild- like wormhole. In order to calculate the radius orbit, let us consider a test particle with rest mass $m$ and four-momentum $p^\mu$ orbiting the equatorial plane. The Lagrangian of this system is written as %\cite{Cataldo:2017}
\begin{align}
\mathcal{L}=\frac{1}{2}mg_{\mu\nu} \frac{dx^\mu}{d\tau} \frac{dx^\nu}{d\tau}.\label{lagrangianorbit}
\end{align} 
Substituting the metric tensor in Eq.\eqref{a4} to Eq.\eqref{lagrangianorbit}, we obtain the Lagrangian in terms of Schwarzschild-like wormhole spacetime,
\begin{align}
\mathcal{L}=\frac{1}{2}m\left\{\Dot{t}^2-F^2(\rho)\Dot{\rho}^2-F^2(\rho)\rho^2\Dot{\theta}^2-F^2(\rho)\rho^2\Dot{\phi}^2\right\},\label{Lagrangianmetrik}
\end{align}
%\begin{eqnarray}
%L=\frac{1}{2}m\left[F^2(\rho)\Dot{\rho}^2+\rho^2F^2(\rho)\Dot{\theta}^2+\rho^2F^2(\rho)\Dot{\phi}^2\right].
%\end{eqnarray}
where the circular orbit is confined to the plane $\theta=\pi/2$. From Eq.\eqref{Lagrangianmetrik}, we calculate $t$- and $\phi$-components of the four-momentum as follows:
%The $t$- and $\phi$-components of the four-momentum in the Schwarzschild-like wormhole spacetime are expressed as follows:
\begin{align}
p_t&=\frac{d\mathcal{L}}{d\Dot{t}}=m\Dot{t}, \label{pt}\\
p_\phi&=\frac{d\mathcal{L}}{d\Dot{\phi}}=-mF^2(\rho)\rho^2 \Dot{\phi}\label{pphi},
\end{align}
where Eqs.\eqref{pt} and \eqref{pphi} satisfy the conserved quantities $p_t=E$ and $p_\phi=-\tilde{L}$, coresponding to the energy and angular momentum of particle, respectively. From these relations, we obtain the equation of motion with respect to the proper time, $\tau$,
\begin{align}
\frac{dt}{d\tau}&=\frac{E}{m},\label{ttau} \\
\frac{d\phi}{d\tau}&=\frac{\tilde{L}}{m\rho^2F^2(\rho)}.\label{phitau}
\end{align}

Combining Eqs.\eqref{ttau} and \eqref{phitau} with the mass-shell condition, $g^{\mu\nu}p_\mu p_\nu=m^2$, leads to the following radial equation
\begin{align}
\left(\frac{d\rho}{d\tau}\right)^2=\frac{1}{F^2(\rho)}\left(\frac{E^2}{m^2}-1-\frac{\tilde{L}^2}{m^2\rho^2 F^2(\rho)}\right).\label{radialmotion}
\end{align}
To solve Eq.\eqref{radialmotion}, we choose the orbit of the test particle in the fixed circular orbit where it satisfies the condition, $d\rho/d\tau=0$. From this condition, we find the angular momentum in the following
\begin{align}
\left(\frac{\tilde{L}}{m}\right)^2=\rho^2 F^2(\rho)\left(\frac{E^2}{m^2}-1\right), \label{lambda}
\end{align}
and the radius orbit of the particle can be obtained by substituting Eq.\eqref{a3} into Eq.\eqref{lambda} as follows:
\begin{align}
  \rho_{\pm}=&r_0  \Bigg\{ \frac{(1 - \gamma)}{2} \Bigg[\left(\frac{\tilde{L}/m}{r_0 \sqrt{(E/m)^2-1}} - \frac{1}{2} \right) \nonumber\\
  & \pm \sqrt{\frac{\tilde{L}/m}{r_0 \sqrt{(E/m)^2 - 1}}\left(\frac{\tilde{L}/m}{r_0 \sqrt{(E/m)^2 - 1}}-1\right)} \Bigg] \Bigg\}^{1/\sqrt{1-\gamma}},\label{radius}
\end{align}
where the radius orbits in Eq.\eqref{radius} exists only for $\tilde{L}/m \ge r_0 \sqrt{(E/m)^2 - 1}$ and $(E/m)^2>1$.

Next, from Eq.\eqref{lambda}, we introduce the effective potential, $V_{\text{eff}}(\rho):=E$, as follows\footnote{This effective potential has been also discussed in Ref.\cite{Santos:2021}.}
\begin{eqnarray}
V_{\textrm{eff}}=m\sqrt{1+\frac{\tilde{L}^2}{m^2\rho^2 F^2(\rho)}}.\label{Veff}
\end{eqnarray}
The critical point $\rho_{c}$ of Eq.\eqref{Veff} can be determined by taking $d^2V_{\textrm{eff}}/d\rho^2<0$. It is of the same form as $\rho_{\text{thr}}$, that is, $\rho_c=\rho_{\text{thr}}=r_0\left(\frac{1-\gamma}{4}\right)^{\frac{1}{\sqrt{1-\gamma}}}$. In this context, $\rho_c$ is defined as the radius of the unstable circular orbit, and considering the orbit that stays at and out of the wormhole throat $\rho \ge r_0$, one obtains $\gamma \le -3$.

\subsection{The Comoving Frame}
This part will introduce a system of parallel plate orbiting in a Schwarzschild-like wormhole spacetime. We assume that only the gravitational influences are considered, neglecting the other external contributions. We move to the rotating coordinates $(t',r',\theta',\phi')$ using the following transformation \cite{Sorge:2020}:
\begin{eqnarray}
dt=dt', \hspace{0,3cm} d\rho=dr', \hspace{0,3cm} d\theta=d\theta', \hspace{0,3cm} d\phi=d\phi'+\Omega dt', \label{transcom}
\end{eqnarray}
where $\Omega=d\phi'/dt$ is the angular velocity of the plate. For $\theta=\pi/2$, the wormhole line element in Eq.\eqref{b0} reads
\begin{eqnarray}
  d s^2=\left[ {1 - r'}^2 \Omega^2 F^2 (r') \right] {d t'}^2 - F^2 (r')
  \left[ {d r'}^2 +{ r'}^2 {d \theta'}^2 +{ r'}^2 {d \phi'}^2 +{ 2 r'}^2
  \Omega d \phi' d t' \right]. \label{metrikrotating}
\end{eqnarray}
The function of $F(r')$ above is easily obtained by applying the transformation Eq.\eqref{transcom} to Eq.\eqref{a3}.
The orbit is set in an equatorial plane, recalling $\theta=\pi/2$, with a counterclockwise rotation.
From Eq.\eqref{metrikrotating}, we can infer the determinant of the metric tensor as follows,
\begin{align}
g'=-F^6(r')r'^4, \label{determinant}
\end{align} and the inverse of the metric tensor is given by
\begin{equation}
g'^{\mu\nu}=
\begin{pmatrix}
1 & 0 & 0 & -\Omega\\
0 & -\frac{1}{F^2(r')} & 0 & 0\\
0 & 0 & -\frac{1}{r'^2 F^2(r')} & 0\\
-\Omega & 0 & 0 & -\frac{1-r'^2 F^2(r')\Omega^2}{r'^2F^2(r')}
\end{pmatrix}\label{inversetensormetrik}.
\end{equation}
The 4-velocity of a comoving observer is defined by a unit tangent vector $u^\mu$ as follows:
\begin{eqnarray}
  u^{\mu}  = (u^{t'}, 0, 0, u^{\phi'}) = u^{t'}(1, 0, 0, \Omega),
\end{eqnarray}
where $u^{t'}$ can be obtained from normalization requirement $g_{\mu\nu}u^\mu u^\nu=1$ as
\begin{align}
  u^{t'} = \frac{1}{\sqrt{1 - {r'}^2 \Omega^2 F^2 ({r'})}},
\end{align}
where the allowed angular velocity satisfies
\begin{eqnarray}
0 \le \Omega < \frac{1- \gamma}{r_0\left( \frac{r'}{r_0} \right)^{\sqrt{(1 - \gamma) }}} \left( 1 + \frac{1 - \gamma}{4 \left( \frac{r'}{r_0}\right)^{\sqrt{(1 - \gamma) }}} \right)^{-2}.
\end{eqnarray}

Next, we introduce an orthonormal tetrad frame describing the local frame in the Casimir system and denote the tetrad coordinates with a caret $(\hat{\tau},\hat{x},\hat{y},\hat{z})$ to avoid ambiguities as follows:
\begin{equation}
    \begin{split}
    \theta^{\hat{\tau}} & =  e_{t'}^{\hat{\tau}} dt', \\
  \theta^{\hat{x}} & =  e_{r'}^{\hat{x}} dr', \\
  \theta^{\hat{y}} & =  e_{\theta'}^{\hat{y}} d\theta',\\
  \theta^{\hat{z}} &=  e_{t'}^{\hat{z}} dt' + \theta_{\phi'}^{\hat{z}} d\phi'.
    \end{split}\label{tetrad}
\end{equation}
The vierbien tetrad is determined by comparing Eq.\eqref{tetrad} with Eq.\eqref{metrikrotating}. By applying the orthonormality principle ${g}'_{\mu\nu} e^{\mu}_{\hat{\alpha}} e^{\nu}_{\hat{\beta}}=\eta_{\hat{\alpha}\hat{\beta}}$, we can write down the inverse vierbein tetrad $e^\mu_{\hat{\alpha}}$ as follows:
\begin{align}
  e^{t'}_{\hat{\tau}} = & u^{t'} = \frac{1}{\sqrt{{1 - r'}^2 \Omega^2 F^2 (r')}}, \nonumber\\
  e^{r'}_{\hat{x}} = & \frac{1}{[F (r')] },\nonumber\\
  e^{\theta'}_{\hat{y}} = & \frac{1}{[r' F (r')] } ,\label{inverstetrad} \\
  e^{t'}_{\hat{z}} = & \frac{r' \Omega F (r')}{\sqrt{{1 - r'}^2 \Omega^2 F^2
  (r')}},\nonumber\\
  e^{\phi'}_{\hat{z}} = & \frac{\sqrt{{1 - r'}^2 \Omega^2 F^2 (r')}}{r' F(r')}. \nonumber 
\end{align}
We note that substituting the inverse vierbein tetrad in Eq.\eqref{inverstetrad} to Eq.\eqref{metrikrotating} leads to the line element equation in a Minkowskian spacetime. Additionally, Eq.\eqref{inverstetrad} is useful when we solve the Klein-Gordon equation in the next section. In the following sections, the primed notation will be omitted for simplicity.

\section{Casimir Energy of the Schwarzschild-like Wormhole}
\label{sec3}
In this section, we derive the Casimir energy of the Schwarzschild-like wormhole following the procedure in Ref.\cite{Sorge:2020}. For this purpose, we first consider the scalar field equation of a massless particle confined inside the plate cavity orbiting in the Schwarzschild-like wormhole. The solution of the scalar field equation satisfies the Dirichlet boundary condition on the plate's surface. In general coordinates, the scalar field equation reads:
\begin{eqnarray}
  \frac{1}{\sqrt{-\hat{g}}} \partial_{\mu} \left[ \sqrt{- \hat{g}} \hat{g}^{\mu \nu} \partial_{\nu} \phi(x) \right] =0, \label{KG}
\end{eqnarray}
where $\hat{g}$ and $\hat{g}^{\mu\nu}$ is given in Eqs.\eqref{determinant} and \eqref{inversetensormetrik}, respectively. In this work, the metric tensor indicates the Minkowski metric tensor. Rewriting Eq.\eqref{KG} in terms of a vierbien tetrad, we obtain the following equation
\begin{align}
  \left[\eta^{\hat{\alpha} \hat{\beta}}   \partial_{\hat{\alpha}}
  \partial_{\hat{\beta}} - \eta^{\hat{x}\hat{x}} \left(\frac{1}{rF(r)}+\frac{3F'(r)}{F^2(r)}-\frac{F'(r)}{F^2(r)}\right) \partial_{\hat{x}} \right]\phi (\hat{x})=0, \label{persKGeksak}
\end{align}
where $\eta^{\hat{\alpha} \hat{\beta}}$ is a Minkowski metric tensor in the comoving frame and $F'(r)$ is defined as the derivative with respect to the radial coordinate $r$. In order to solve Eq.\eqref{persKGeksak}, we assume that the plates orbit around the Schwarzschild-like wormhole at a fixed radius orbit. Consequently, $F(r)$ is considered as a constant and $F'(r)$ goes to zero. Therefore, Eq.\eqref{persKGeksak} can now be written as follows:
\begin{align}
  \left[\eta^{\hat{\alpha} \hat{\beta}}   \partial_{\hat{\alpha}}
  \partial_{\hat{\beta}} - \frac{2}{{\hat{r}} } \partial_{\hat{x}}
  \right]\phi (\hat{x})=0, \label{KGlokal}
\end{align}
with $\hat{r}\equiv rF(r)$. 

Next, we apply the Dirichlet boundary conditions in Eq.\eqref{dirichlet} on the plate's surface of the Casimir cavity. By using the separation method of variables, we obtain the solution of Eq.\eqref{KGlokal} as follows:
\begin{eqnarray}
  \phi (\hat{\tau}, \hat{x}, \hat{y}, \hat{z}) = N_n e^{- i \omega_n \hat{\tau}} e^{ik_{\hat{y}} \hat{y}} e^{-
  \frac{\hat{x}}{\hat{r}} } e^{ik_{\hat{x}} \sqrt{\left( 1 - \frac{1}{\hat{r}^2 k^2_{\hat{x}}} \right)}  \hat{x}} \sin \left( \frac{n \pi}{L} \hat{z}\right), \label{solkg}
\end{eqnarray}
and the eigen-frequency is given by
\begin{eqnarray}
  \omega_n^2 = \textbf{k}^2_{_{\bot}} + \left( \frac{n \pi}{L} \right)^2, \label{eigenfrek}
\end{eqnarray}
where $\textbf{k}=(k_{\hat{x}},k_{\hat{y}},k_{\hat{z}})=(\textbf{k}_\bot,n\pi/L)$ is the momenta of the propagation modes parallel to the plates. Since every point inside the Casimir cavity satisfies $\left\{\hat{x},\hat{y}\right\}<L$ and Eq.\eqref{model}, we have $\frac{\hat{x}}{\hat{r}}\ll 1$, implying that the field modes reduce to their Minkowskian form. Note that the eigen-frequency is obtained by applying the Dirichlet boundary condition to the plate placed on the \textit{$\hat{z}$}-axis. $N_n$ is the normalization constant and it can be obtained from the orthonormality requirement \cite{Birrel:1984}:
\begin{eqnarray}
  \langle \phi_n (\textbf{k}), \phi_m (\textbf{k}')\rangle = i \int_\Sigma  [(\partial_\mu  \phi_n)\phi^{*}_m - \phi_n (\partial_\mu \phi^{*}_m)] n^\mu \sqrt{g_\Sigma} d\Sigma,\label{ortho}
\end{eqnarray}
in which $\Sigma$ shows the spacelike Cauchy surface $d\Sigma=d\hat{x} d\hat{y} d\hat{z}$,  $g_\Sigma=-\text{det}(g_{\mu\nu})/g_{\hat{\tau}\hat{\tau}}$ is the determinant of the induced 3D-metric on a given timelike oriented hypersurface (the cavity), and $n^\mu$ is a timelike unit vector: $n^\mu=\left(1,0,0,-\Omega\right)$. In particular, Eq.\eqref{ortho} in the tetrad frame has the following form,
\begin{eqnarray}
  \langle \phi_n (\textbf{k}), \phi_m (\textbf{k}')\rangle = i \int_S  [(\partial_a  \phi_n)\phi^{*}_m - \phi_n (\partial_a \phi^{*}_m)] n^a \sqrt{g_S} d\hat{x} d\hat{y} d\hat{z}.\label{normtetrad}
\end{eqnarray}
%where $n^a=e^a_\mu n^\mu$.
Meanwhile, the left hand side of the above equation can be written as
\begin{eqnarray}
  \langle \phi_n (\textbf{k}), \phi_m (\textbf{k}')\rangle = \delta^2(\textbf{k}_\bot-\textbf{k}'_\bot) \delta_{mn}. \label{norm}
\end{eqnarray}
Equalizing Eqs.\eqref{normtetrad} and  \eqref{norm}, we obtain
\begin{eqnarray}
  N_n = \left( \frac{ \sqrt{{1 - r}^2 \Omega^2 F^2 (r)}}{4 \pi^2
  \omega_n L} \right)^{1 / 2}.
\end{eqnarray}

Now, we evaluate the Casimir energy density for the scalar field inside the orbiting cavity. The time component of the energy-momentum tensor of the scalar field is given by
\begin{align}
  T_{\hat{\tau} \hat{\tau}} =\partial_{\hat{\tau}} \phi_n \partial_{\hat{\tau}} \phi^{\ast}_n -
  \frac{1}{2} \eta^{ab} \partial_a \phi_n \partial_b
  \phi^{\ast}_n. \label{TensorEM0}
\end{align}
By using Eq.\eqref{solkg}, we can compute Eq.\eqref{TensorEM0} as follows,
\begin{eqnarray}
    T_{\hat{\tau} \hat{\tau}} =\frac{1}{2} \left[ \omega^2_n \phi_{n, \textbf{k}_{_{\bot}}} \phi^{\ast}_{n,
\textbf{k}_{_{\bot}}} + \textbf{k}^2_{\bot} \phi_{n, \textbf{k}_{_{\bot}}} \phi^{\ast}_{n, \textbf{k}_{_{\bot}}} +{N^2_n  \left( \frac{n \pi}{L} \right)^2 \cos^2 \left( \frac{n
\pi}{L} \hat{z} \right)} \right].\label{TensorEM}
\end{eqnarray}
%Adopting the procedure in Ref.\cite{Sorge:2020},
The mean vacuum energy density can be calculated as
\begin{eqnarray}
  \Bar{\varepsilon}_{\text{vac}} = \frac{1}{V_p} \int_{\Sigma} d^3 \hat{x} \underset{n}{\sum} 
  \int dk_{\hat{x}} dk_{\hat{y}} T_{\hat{\tau} \hat{\tau}} [\phi_n, \phi^{\ast}_n], \label{evakum}
\end{eqnarray}
where $V_p=V\sqrt{-g_\Sigma}$ is the proper cavity volume. Substituting Eq.\eqref{TensorEM} into Eq.\eqref{evakum} yields
\begin{eqnarray}
    \Bar{\varepsilon}_{\text{vac}} =\frac{\sqrt{{1 - r}^2 \Omega^2 F^2 (r)}}{8 \pi^2 L} \underset{n}{\sum} \int d^2 \textbf{k}_{\bot}  \sqrt{\textbf{k}^2_{\bot} + \left( \frac{n \pi}{L} \right)^2 }.\label{ecas0}
\end{eqnarray}
Integration in Eq.\eqref{ecas0} can be performed using the Schwinger proper-time method, while summing over $n$ is performed utilizing the Riemann zeta function regularization procedure \cite{sorge:2014}, $\zeta(s)= \sum_{n=1}^\infty n^{-s}$. By doing so, Eq.\eqref{ecas0} leads to the following,
\begin{eqnarray}
  \Bar{\varepsilon}_{\text{Cas}}=\sqrt{{1 - r}^2 \Omega^2 F^2 (r)}  \varepsilon_{\text{Cas}}^{(0)},\label{ecas}
\end{eqnarray}
where $\varepsilon_{\text{Cas}}^{(0)}=-\frac{\pi^2}{1440 L^4}$ is the standard Casimir energy density in flat spacetime. We note that the Casimir energy density in Schwarzschild-like spacetime decreases relative to that of the flat spacetime. On top of that, its value depends on the angular velocity of the plates $\Omega$, the orbital radius of the Casimir system $r$, the distance of plates $L$, and the conformal factor $F(r)$ which determines the background of the Schwarzschild-like wormhole.

\section{Thermal Casimir Effect}
\label{sec4}
In this section, we investigate the finite-temperature implications to the Casimir cavity, following the approach developed in Ref.\cite{Geyer:2008}. To derive the Casimir energy density, we employ the zeta function regularization procedure. In deriving this Casimir energy, we will assume that the temperature is uniform in spacetime. Additionally, we also derive the Casimir free energy, entropy, internal energy, and heat energy capacity at constant volume.

\subsection{Thermal Correction of Casimir Free Energy}
In this calculation, the effect of temperature is considered from a reference frame shared by an observer moving through the cavity of the Casimir system. The Casimir free energy associated with the vacuum in a finite volume $V$ is given by 
\begin{eqnarray}
    \mathcal{F}=E_{\text{Cas},0}+\Delta_T \mathcal{F}_0,\label{casfree}
\end{eqnarray}
where $E_{\text{Cas},0}=V\Bar{\varepsilon}_{\text{Cas}}$ is the renormalized Casimir free energy at zero temperature associated with a volume $V$ and $\Delta_T \mathcal{F}_0$ is the thermal corrections of the Casimir energy in the Matsubara formalism. Explicitly, these thermal corrections can be expressed as follows,
\begin{eqnarray}
\Delta_T \mathcal{F}_0=T \sum_j \ln{\left(1-e^{-\beta \omega_j}\right)},
\end{eqnarray}
where $\beta=1/T$. Specifically, in our case, the above formula is written as follows:
\begin{eqnarray}
  \Delta_T \mathcal{F}_0  & = & \frac{1}{L} \underset{n = 1}{\overset{\infty}{\sum}} \underset{-
  \infty}{\overset{\infty}{\int}} \frac{dk_x dk_y}{(2 \pi)^2}
  \int_V dx dy dz \sqrt{g_\Sigma} T \ln (1 - e^{- \omega_n / T}),\nonumber\\
  & = & \frac{S}{\beta}\underset{n = 1}{\overset{\infty}{\sum}} \underset{-
  \infty}{\overset{\infty}{\int}} \frac{dk_x dk_y}{(2 \pi)^2}
   \ln (1 - e^{- \beta\omega_n}),\label{korek}
\end{eqnarray}
where $\omega_n$ is defined in Eq.\eqref{eigenfrek}. The volume integration of Eq.\eqref{korek} is $\int_V dx dy dz \sqrt{g_\Sigma}=SL=V$ in which the volume of the cavity is measured from the comoving observer. The momentum $\textbf{k}$ has two continuous components $(k_x,k_y)$ parallel to the plates with a relation\footnote{As a consequence, we will have $dk_x dk_y=d^2\textbf{k}_\bot=2\pi \textbf{k}_\bot d\textbf{k}_\bot $ in the polar coordinate.}, $\textbf{k}^2_\bot=k^2_x+k^2_y$ and one discrete momentum $k_z$ perpendicular to plate. By doing so, we can recast Eq.\eqref{korek} into the following
\begin{eqnarray}
\Delta_T \mathcal{F}_0 =\frac{S}{2 \pi \beta} \underset{n = 1}{\overset{\infty}{\sum}} \underset{0}{\overset{\infty}{\int}} \textbf{k}_{\bot} \tmop{d\textbf{k}}_{\bot} \ln \left( 1 - e^{-
\beta \sqrt{\textbf{k}^2_{\bot} + \left( \frac{n \pi}{L} \right)^2}} \right). \label{correct4}
\end{eqnarray}

To perform the integration over momentum in Eq.\eqref{correct4}, we need to expand the logarithmic function in a power series and obtain the following,
\begin{eqnarray}
\Delta_T \mathcal{F}_0 =- \frac{S }{2 \pi \beta} \overset{\infty}{\underset{n, m = 1}{\sum}}
\frac{1}{m} \overset{\infty}{\underset{0}{\int}} d\textbf{k}_{\bot} \textbf{k}_{\bot}
e^{- m \beta \sqrt{\textbf{k}^2_{\bot} + \left( \frac{n \pi}{L} \right)^2}} \label{correct}.
\end{eqnarray}
Next, by introducing $z\equiv \beta \sqrt{\textbf{k}^2_{\bot} + \left( \frac{n \pi}{L} \right)^2}$ and performing the summation over $n$, the thermal correction of the Casimir free energy in Eq.\eqref{correct} is written in terms of a hyperbolic function as follows,
\begin{eqnarray}
\Delta_T \mathcal{F}_0 =- \frac{S }{32 \pi L^3} \overset{\infty}{\underset{m = 1}{\sum}}  \left[
\frac{\pi}{(\tilde{\beta} m)^2 } \frac{1}{\sinh^2 (m \tilde{\beta} \pi)} +
\frac{\coth (m \tilde{\beta} \pi)}{(\tilde{\beta} m)^3} \right] + \frac{S  \xi
(3)}{32 \pi L^3 \tilde{\beta}^3},\label{nonapprok}
\end{eqnarray}
where $\zeta(3)=\sum_{m=1}^\infty=1/m^3$ is the Riemann zeta function and $\tilde{\beta}\equiv \beta/2L$. We note that the above expression has the divergent property in the certain limit of $\tilde{\beta}$ (or temperature). In the next subsection, we will discuss a zeta function regularization method to tackle this matter and consider some limits of $\tilde{\beta}$. 

\subsection{Regularization and Renormalization of Casimir Thermal Correction}
In fact, the thermal correction of the Casimir free energy in Eq.\eqref{nonapprok} contains the asymptotic behavior at high temperature or large separations $\tilde{\beta} \ll 1$. To regularize it, we evaluate the summation over $m$ by using the zeta function regularization procedure and perform the series expansion for $\tilde{\beta} \to 0$. In particular, we have used two approximate forms, namely $\sinh^2{(m\tilde{\beta}\pi)}\simeq (m\tilde{\beta}\pi)^2$ and $\coth{(m\tilde{\beta}\pi)}\simeq 1/(m\tilde{\beta}\pi)$. As a result, we obtain the following expression,
\begin{eqnarray}
    \Delta_T \mathcal{F}_0 \simeq  - \frac{V\pi^2}{1440  (\tilde{\beta}L)^4} - \frac{V \pi^2}{720 L^4} + \frac{S  \xi (3)}{32 \pi (\tilde{\beta}L)^3}, \label{approthermal}
\end{eqnarray}
where we have used $\zeta(4)=\pi^4/90$. 

We note that the correction of the Casimir free energy in Eq.\eqref{approthermal} contains divergent parts due to $\tilde{\beta}$, particularly the first and second terms. In solving this issue, we will adopt a procedure presented in Ref.\cite{Geyer:2008}. This process can be realized by subtracting the asymptotic limit term at high temperature in Eq.\eqref{approthermal} from Eq.\eqref{nonapprok}. Consequently, we will obtain the terms such as, $\kappa_0(1/\tilde{\beta}^2), \kappa_1(1/\tilde{\beta}^3),$ and $\kappa_2(1/\tilde{\beta}^4)$, in which $\kappa_i$ $(i=0,1,2)$ parameters are geometric coefficients related to the plate cavity and they are expressed as heat kernel coefficients \cite{Bordag:2009}. Therefore, the thermal correction of the renormalized Casimir free energy is obtained as follows\footnote{In deriving the thermal correction of the renormalized Casimir free energy, we have ignored the second term of Eq.\eqref{approthermal} since it does not relevant to this analysis.}:
\begin{eqnarray}
    \Delta_T \mathcal{F}^{\text{ren}} =- \frac{S }{32 \pi L^3} \overset{\infty}{\underset{m = 1}{\sum}}  \left[\frac{\pi}{(\tilde{\beta} m)^2 } \frac{1}{\sinh^2 (m \tilde{\beta} \pi)}+\frac{\coth (m \tilde{\beta} \pi)}{(\tilde{\beta} m)^3} \right] +\frac{V  \pi^2}{1440 L^4 \tilde{\beta}^4}. \label{renthermal}
\end{eqnarray}
The last term of Eq.\eqref{renthermal} is actually $-V f_{bb}(\tilde{\beta})$ in which $f_{bb}(\tilde{\beta})=-\frac{\pi^2}{1440 L^4 \tilde{\beta}^4}$ is the free energy density of the black-body radiation analogous to the Minkowski spacetime. Then, the renormalized Casimir free energy, $\mathcal{F^\text{ren}}$, can be written as
\begin{align}
\mathcal{F}^{\text{ren}} %&= E_{\text{Cas,0}}+\Delta_T \mathcal{F}^{\text{ren}} \nonumber\\
&=E_{\text{Cas,0}}-\frac{S }{32 \pi L^3} \overset{\infty}{\underset{m = 1}{\sum}}  \left[
\frac{\pi}{(\tilde{\beta} m)^2 } \frac{1}{\sinh^2 (m \tilde{\beta} \pi)} +
\frac{\coth (m \tilde{\beta} \pi)}{(\tilde{\beta} m)^3} \right] - V f_{bb}(\tilde{\beta})\label{fren}.
\end{align}
From this result, we can see that thermal correction of the renormalized Casimir free energy depend on the temperature and the plate's distance, but it does not depend on the background of the spacetime. %\textcolor{red}{As shown in Eq.\eqref{fren}, we find that the Casimir energy decreases at finite temperature. This result is significant because the thermal effects can modify the wormhole throat geometry: the size and stability of the throat \cite{hotcas:2025}.}

Let us compare our thermal correction of the Casimir free energy with that of Ref. \cite{zhang:2015}. Their expression contains a proper length $L_p$ defined as $L_p\equiv L (\sqrt{\Delta} / r)  C(\Omega)$ with $\Omega\equiv d \phi / d t$ being an angular velocity and a proper surface of the plate $S_p$ defined as  $S_p\equiv S\left(\sqrt{\Delta} / r \right)^{- 1}$ 
\footnote{$\Delta = r^2 + a^2 - 2 M r$ is a prefactor that is defined in the Kerr spacetime metric \cite{zhang:2015}.}. Those two quantities will determine the geometry of plates. In contrast, our expression for the thermal corrections does not have such scaled quantity as the spacetime background of our Casimir cavity is locally flat (Minkowskian) and the determinant of the metric is unity. It is also important to mention that the analysis of both results is done in the comoving observer and our result is indeed a special case of Kerr spacetime by setting a quantity $C (\Omega) = 1$ \cite{zhang:2015}.

\begin{figure}
    \centering
    \includegraphics[width=0.6\linewidth]{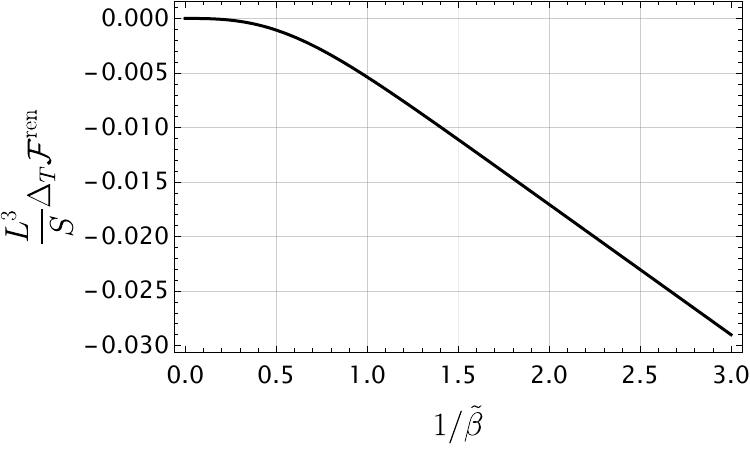}
    \caption{Plot of $\frac{L^3}{S}\Delta_T \mathcal{F}^{\text{ren}}$ as a function of dimensionless parameter, $1/\tilde{\beta}$. We perform the summation over $m$ up to 1000.}
    \label{figfree}
\end{figure}

In figure \ref{figfree}, we show the thermal correction of the renormalized Casimir free energy as a function of the dimensionless parameter $1/\tilde{\beta}$ (recalling $1/\tilde{\beta}=2LT$). For the numeric calculation, we perform the summation over $m$ up to 1000. This is because for larger values of $m$, the term inside the brackets of Eq.\eqref{renthermal} tends to zero; thus its contribution is negligible. We find that the thermal correction of the renormalized Casimir free energy gradually decreases as the dimensionless temperature, $1/\tilde{\beta}$, increases. In particular, in the range of $1/\tilde{\beta}<0.3$, the dominant contribution comes from the second term of Eq.\eqref{renthermal} in the order of $\tilde{\beta}^{-3}$. Meanwhile, for the range of $1/\tilde{\beta}\gtrsim 0.3$, the contribution is proportional to a term with the order of $\tilde{\beta}^{-4}$. 

%In particular, at the beginning: $0<L<0.05$ (black-solid), $0<L<0.2$ (blue-dotted), and  $0<L<0.5$ (red-dashed), the curves increase monotonically as the plate distance, $L$, increase. This is because the summation part in Eq.\eqref{renthermal} is more dominant compared to that of the last term. In the range: $0L\ge 0.05$ (black-solid), $L\ge0.2$ (blue-dotted), and  $L\ge0.5$ (red-dashed), each of the curves approach zero because the difference of the absolute value from the summation part and the last term of Eq.\eqref{renthermal} leading to a very tiny value.}

\begin{figure}[ht!]
    \centering
    \includegraphics[width=0.6\linewidth]{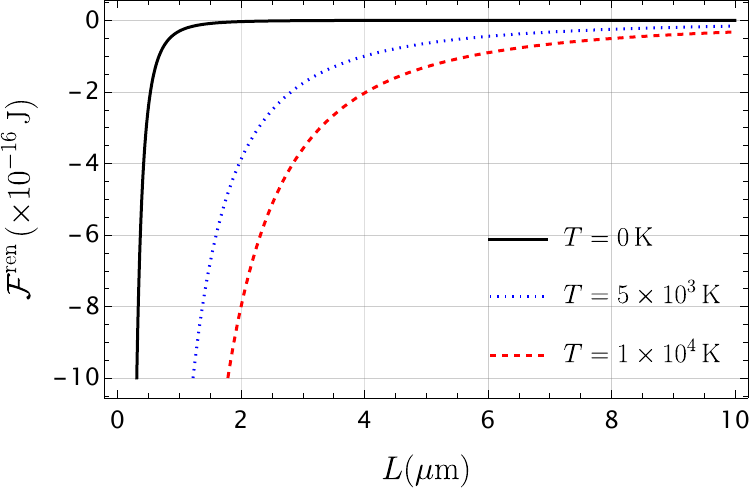}
    \caption{Plot of the renormalized Casimir free energy, $\mathcal{F}^{\text{ren}}$, as a function of the plate's distance, $L$. All quantities are expressed in SI units. The black-solid, red-dashed, and blue-dotted lines correspond to the renormalized Casimir free energy at $0$ K, $5\times 10^{3}$ K, and $1\times 10^{4}$ K, respectively. For each temperature, we perform the summation over $m$ up to 1000.}
    \label{FrentoL}
\end{figure}

We next study the renormalized Casimir free energy, $\mathcal{F}^{\text{ren}}$, as a function of the plate's distance, $L$, at finite temperature. For this purpose, we should write all quantities in Eq.\eqref{fren} in SI units. In these units, each term in the renormalized Casimir free energy expression should be multiplied by a factor $\hbar c$. On the other hand, the parameter $\tilde{\beta}$ should be written as $\tilde{\beta}=\frac{\hbar c}{2k_B TL}$, with $k_B$ being a Boltzmann constant. For the numerical calculation, we set $\gamma=-3$, corresponding to an orbit located at the wormhole throat; the plate's area $S$ in millimeter units; and $r=2.95\times 10^4$ m corresponding to the Schwarzschild radius of a stellar-mass black hole. With all these configurations, we can constrain the allowed temperature region of Eq.\eqref{fren}. Specifically, within the high temperature limit, we compute the ratio $|V f_{bb} (\tilde{\beta})/E_{\text{Cas,0}}|\le 1$ and obtain the allowed temperature in the range $0 \le T\le 10^5$ K\footnote{In deriving this allowed temperature region, one notes that the last term of Eq.\eqref{fren} could be dominant at the high temperature limit. Therefore, one should take an inequality $|V f_{bb}(\tilde{\beta})/E_{\text{Cas,0}}|\le 1$, keeping the plate's distance in millimeter units. If one takes a temperature beyond the upper bound temperature values, the $V f_{bb}(\tilde{\beta})$  term will be dominant, meaning that the thermal correction is greater than $E_{\text{Cas,0}}$ and there will be a transition from the attractive energy/force to the repulsive ones. In any case, this situation is unreasonable.}. In figure \ref{FrentoL}, we show the renormalized thermal Casimir free energy as a function of the plate's distance, where the black-solid, red-dashed, and blue-dotted lines correspond to the renormalized Casimir free energy at 0 K, $5\times 10^3$ K, and $1\times 10^4$ K, respectively. From this figure, we find that as the temperature increases, the magnitude of the renormalized Casimir free energy increases and vice versa. Our result also agrees with that of Ref.\cite{sasanpour:2023}.

Now, we compute the renormalized Casimir entropy as defined in the following,
\begin{eqnarray}
    \mathcal{S}^{\text{ren}}=-\frac{\partial\mathcal{F}^{\text{ren}}_{\text{tot}}}{\partial T}=- \frac{d \mathcal{F}^{\text{ren}}_{\text{tot}}}{d  \tilde{\beta}}  \frac{d \tilde{\beta}}{dT}.\label{entropi0}
\end{eqnarray}
By recalling $\tilde{\beta}=1/2LT$ and substituting Eq.\eqref{fren} to Eq.\eqref{entropi0}, one obtains as follow:
\begin{align}
    \mathcal{S}^{\text{ren}}=\frac{3S }{16 \pi L^2} \left\{ \overset{\infty}{\underset{m =
1}{\sum}}  \left[ \frac{\pi}{\tilde{\beta} m^2 \sinh^2 (m \tilde{\beta} \pi)}
+ \frac{2 \pi^2 \coth (m \tilde{\beta} \pi)}{3 m \sinh^2 (m \tilde{\beta}
\pi)} + \frac{\coth (m \tilde{\beta} \pi)}{\tilde{\beta}^2 m^3 } \right] - \frac{4 \pi^3}{135 \tilde{\beta}^3 } \right\}. \label{sren}
\end{align}
The renormalized Casimir entropy in Eq.\eqref{sren} is equal to the thermal correction of the renormalized Casimir entropy,  $\mathcal{S}^{\text{ren}}=\Delta_T \mathcal{S}^{\text{ren}}$, because $E_{\text{Cas},0}$ does not depend on the temperature. 

Next, we calculate the renormalized Casimir internal energy, which is defined as follows:
\begin{align}
  \mathcal{U}^{\text{ren}} (T)=- T^2 \frac{\partial}{\partial T} \left(
  \frac{\mathcal{F}^{\text{ren}}_{\text{tot}}}{T} \right)=-\frac{1}{2L\tilde{\beta}^2}\frac{d}{d\tilde{\beta}}\left(\tilde{\beta} \mathcal{F}^{\text{ren}}_{\text{tot}}\right)\frac{d \tilde{\beta}}{dT}. \label{energiinternal0}
\end{align}
By using Eqs.\eqref{fren} and \eqref{energiinternal0}, one obtains
\begin{align}
\mathcal{U}^{\text{ren}}= E_{\text{Cas,0}} + \frac{S }{16 \pi L^3}{\left\{
\overset{\infty}{\underset{m = 1}{\sum}} \left[ 
\frac{\pi}{\tilde{\beta}^2 m^2 \sinh^2 (m \tilde{\beta} \pi)} + \frac{\pi^2
\coth (m \tilde{\beta} \pi)}{\tilde{\beta} m \sinh^2 (m \tilde{\beta} \pi)} +
{{\frac{\coth (m \tilde{\beta} \pi)}{
\tilde{\beta}^3 m^3 }}} \right] - \frac{\pi^3}{30
\tilde{\beta}^4} \right\}}. \label{uren}
\end{align}
We note that the second term of the above expression is the thermal correction of the renormalized  Casimir internal energy $\Delta_T \mathcal{U}^{\text{ren}}$, which depends on the temperature and plate's size.

Moreover, we also compute the renormalized Casimir heat energy capacity at a constant volume as defined in the following
\begin{align}
\mathcal{C}_V^{\text{ren}}=\left(\frac{\partial \mathcal{U}^{\text{ren}}}{\partial T}\right)_V=\left(\frac{d\mathcal{U}^{\text{ren}}}{d\tilde{\beta}}\right)_V \frac{d\tilde{\beta}}{dT}. \label{C_v}
\end{align}
Substituting Eq.\eqref{uren} to the Eq.\eqref{C_v}, we obtain
\begin{align}
\mathcal{C}_V^{\text{ren}}=&\frac{S
k_{\beta}}{8 \pi L^2} \Bigg[ \overset{\infty}{\underset{m =
1}{\sum}} \Bigg\{\frac{3
\pi}{\tilde{\beta} m^2 \sinh^2 (m \tilde{\beta} \pi)}+ \frac{3 \coth (m
\tilde{\beta} \pi)}{\tilde{\beta}^2 m^3 }+\frac{3 \pi^2 \coth (m \tilde{\beta} \pi)}{m \sinh^2 (m \tilde{\beta} \pi)} + \frac{\tilde{\beta} \pi^3}{\sinh^4 (m\tilde{\beta} \pi)}+ \nonumber\\
& \frac{2 \pi^3 \tilde{\beta} \coth^2 (m \tilde{\beta}\pi)}{\sinh^2 (m \tilde{\beta} \pi)}\Bigg\} - \frac{4 \pi^3}{30\tilde{\beta}^3} \Bigg], \label{Cvren}
\end{align}
where the renormalized Casimir heat energy capacity at constant volume is equal to the thermal correction of the renormalized Casimir heat energy capacity, $\mathcal{C}_V^{\text{ren}}=\Delta_T \mathcal{C}_V^{\text{ren}}$.

\begin{figure}[t]
    \centering
    \includegraphics[width=0.6\linewidth]{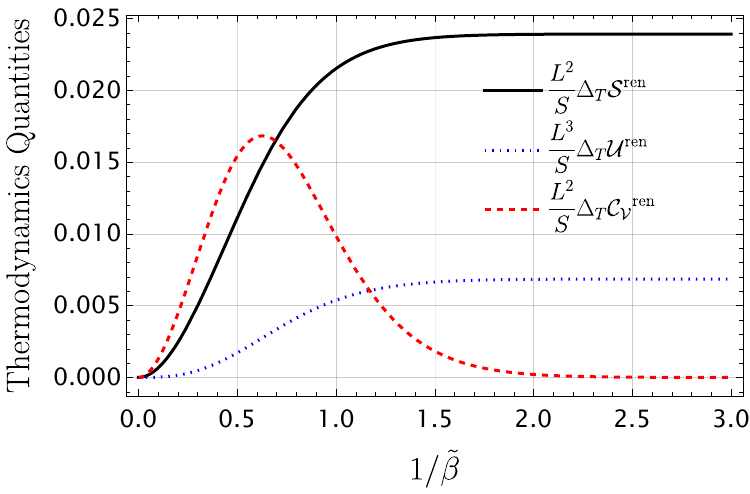}
    \caption{Plot of thermodynamics quantities: $\frac{L^2}{S}\Delta_T \mathcal{S}^{\text{ren}}$, $\frac{L^3}{S}\Delta_T \mathcal{U}^{\text{ren}}$ and $\frac{L^2}{S}\Delta_T \mathcal{C_V}^{\text{ren}}$ as a function of dimensionless parameter, $1/\tilde{\beta}$. The black-solid, blue-dotted, and red-dashed lines correspond to the thermal correction of the renormalized Casimir entropy, internal energy, and heat capacity at constant volume, respectively. For each quantity, we perform the summation over $m$ up to 1000.}
    \label{fig3}
\end{figure}
In figure \ref{fig3}, we plot thermodynamic quantities of the renormalized Casimir system as a function of the dimensionless parameter $1/\tilde{\beta}$. The black-solid, blue-dotted, and red-dashed lines represent the thermal correction of the renormalized Casimir entropy, internal energy, and heat capacity at a constant volume, respectively. For each quantity, we have also performed the summation over $m$ up to 1000. The thermal corrections of the renormalized Casimir entropy and internal energy exhibit a similar behavior. In particular, at the beginning $(0<1/\tilde{\beta}<1.5)$, the black-solid line has a higher rise compared to that of the red-dashed line. Within this interval, both curves increase monotonically as the dimensionless temperature  $1/\tilde{\beta}$ increases. This is because the summation part in Eqs.\eqref{sren} and \eqref{uren} is more dominant compared to that of the last term. At $1/\tilde{\beta}\gtrsim 1.5$, both curves approach finite asymptotic values because the absolute value of the summation part in Eqs.\eqref{sren} and \eqref{uren} is slightly different compared to that of the last term. Next, we consider the thermal correction of the renormalized Casimir heat capacity at a constant volume that corresponds to the red-dashed line. At the beginning ($1/\tilde{\beta}<0.6$), the curve increases monotonically with increasing $1/\tilde{\beta}$ and exhibits a maximum at $1/\tilde{\beta}\simeq 0.6$. Specifically, in this interval, the contribution is dominated mainly by the second term of the summation part compared to other terms in Eq.\eqref{Cvren}. It turns out that, there was a transition, in which the curve gradually decreases in the range of $0.6<1/\tilde{\beta}<2$ as $1/\tilde{\beta}$ increases due to the last term of Eq.\eqref{Cvren}. In the range of $1/\tilde{\beta}\gtrsim2$, the curve approaches zero. This can be understood if we take the absolute value difference of the summation part with the last term of Eq.\eqref{Cvren}, leading to a very tiny value. To conclude this subsection, it is also important to mention that the behavior of the heat capacity at constant volume is consistent with the fundamental laws of thermodynamics.

\subsection{Low Temperature Limit} \label{seclowtemp}

In this section, we study the asymptotic properties of thermal corrections for Casimir thermodynamic quantities at very low temperature $1 /
\tilde{\beta} \to 0$. For this purpose, we write hyperbolic functions in Eq.\eqref{fren} in terms of exponential form. In order to do so, we replace the hyperbolic functions with $1 / (1 - e^{- 2 \pi m \tilde{\beta}}) \simeq 1 + e^{- 2 \pi m\tilde{\beta}}$ and keep their expansion up to $\mathcal{O} (e^{- 2 \pi m\tilde{\beta}})$. In particular, we consider thermal corrections beyond the influence of power-law temperature.

In the following, we compute the thermal correction of the renormalized Casimir free energy in this limit. We first focus on this quantity as the other thermodynamic quantities can be deduced on the basis of this expression. Within this limit, the renormalized Casimir free energy is computed as follows,
\begin{align}
\mathcal{F}_{\text{low}}^{\text{ren}}\simeq & E_{\text{Cas,0}} - \frac{S  \zeta (3)}{32 \pi L^3 \tilde{\beta}^3} +\frac{V \pi^2}{1440 L^4 \tilde{\beta}^4 } - \frac{S}{8 L^3}\overset{\infty}{\underset{m =1}{\sum}} \left(1+\frac{1}{2 \pi \tilde{\beta}m}\right) \frac{e^{- 2 m \pi \tilde{\beta}}}{\tilde{\beta}^2 m^2}.\label{corlowtempfreeener}
\end{align}
We should remind that the first term of the above equation is the zeroth-order temperature of the Casimir energy that originates from the vacuum fluctuations due to boundary conditions, while the second term gives the leading thermal correction localized at the boundary and represents modified low frequency Matsubara modes. The next-to-leading power-law term, proportional to $\pi^2 / \tilde{\beta}^4$, corresponds to a bulk thermal contribution and it shows a characteristic of the blackbody radiation energy density confined within the cavity volume $V$, becoming significant only at the higher temperature. Beyond these power-law terms, all remaining contributions are exponentially suppressed by a factor $e^{- 2 \pi m \tilde{\beta}}$. All these thermal correction terms favor the explanation of figure \ref{figfree}, particularly, the behavior of the curve at beginning regime ($1 / \tilde{\beta} < 0.3$).
 
To conclude this section, we write the remaining other thermodynamic quantities at low temperature limit which are given by,
\begin{align}
\mathcal{S}_{\text{low}}^{\tmop{ren}} \simeq & \frac{3S\zeta (3)}{16 \pi L^2 \tilde{\beta}^2} - \frac{S\pi^3}{180 L^2 \tilde{\beta}^3} + \frac{\pi S}{2 L^2} \overset{\infty}{\underset{m = 1}{\sum}} \left(1+ \frac{3}{2\pi\tilde{\beta}m} + \frac{3}{4 \pi^2 \tilde{\beta}^2 m^2}\right) \frac{e^{- 2 m \tilde{\beta} \pi}}{m},\\
\mathcal{U}_{\text{low}}^{\text{ren}}\simeq & E_{\text{Cas,0}} + \frac{S  \zeta (3)}{16 \pi L^3 \tilde{\beta}^3} - \frac{S \pi^2}{480 L^3 \tilde{\beta}^4} +\frac{S \pi}{4 L^3}\overset{\infty}{\underset{m = 1}{\sum}} \left(1 +\frac{1}{\pi \tilde{\beta}} m +\frac{1}{2\pi^2 \tilde{\beta}^2 m^2}\right) \frac{e^{- 2 m \tilde{\beta} \pi}}{\tilde{\beta} m},\\
\mathcal{C}_{V,\text{low}}^{\text{ren}}\simeq & \frac{3 S \xi (3)}{8 \pi L^2 \tilde{\beta}^2}-\frac{\pi^2 S}{60 L^2 \tilde{\beta}^3} +\frac{3S}{2 L^2} \overset{\infty}{\underset{m = 1}{\sum}} \left(1 + \frac{1}{2\pi \tilde{\beta} m}  + \pi \tilde{\beta} m + \frac{2\pi^2 \tilde{\beta}^2m^2}{3}\right) \frac{e^{-2m\tilde{\beta} \pi}}{\tilde{\beta} m^2},
\end{align}
where $\mathcal{S}_{\text{low}}^{\tmop{ren}}=\Delta_T \mathcal{S}_{\text{low}}^{\tmop{ren}}$ and $\mathcal{C}_{V,\text{low}}^{\text{ren}}=\Delta_T \mathcal{C}_{V,\text{low}}^{\text{ren}}$. For the case that the temperature goes to zero, the Casimir internal energy $\mathcal{U}^{\text{ren}}$ goes to $E_{\text{Cas,0}}$ and the Casimir entropy $\mathcal{S}^{\text{ren}}$ is zero, which is suitable with the third law of thermodynamics, according to the Nernst heat theorem \cite{Bezerra:2011, Landau:2013}.

\section{Summary}
\label{sec5}
In this work, we have investigated the finite-temperature Casimir effect for a massless scalar field confined between two parallel plates orbiting a static and zero-tidal Schwarzschild-like wormhole spacetime. %\textcolor{red}{We extend the background in the previous Casimir energy analysis of Ref.\cite{Sorge:2020} to a Schwarzschild-like wormhole spacetime and investigate its thermal effect following the formalism presented in Ref.\cite{Geyer:2008}.}
The calculation of the vacuum energy density is performed in the comoving frame by determining an orthonormal tetrad frame. The tetrad (vierbein) transformation is applied to express the scalar field equation in a locally comoving frame. By imposing Dirichlet boundary conditions on the plates, we obtain the corresponding mode of a scalar field with a discretized momentum. The vacuum energy density is determined from the time component of the energy-momentum tensor of the scalar field. In fact, the vacuum energy density is divergent. Using the Schwinger proper-time method and the Riemann zeta regularization procedure, we obtain the Casimir energy density in a Schwarzschild-like spacetime that decreases relative to the Casimir energy in a flat spacetime, as shown in Eq.\eqref{ecas}. 

The finite-temperature of the Casimir energy density is derived using the Matsubara formalism, leading to the thermal correction of the Casimir free energy. The summation over $n$ of the thermal correction to the Casimir free energy in terms of a hyperbolic function is divergent and we solve it using suitable regularization and renormalization procedures. 
A central result is that, after renormalization, these thermal corrections remain unaffected by the wormhole spacetime background, depending only on the temperature and the plate's distance. This behavior can be understood from the fact that our analysis is performed in a comoving frame, which is locally equivalent to flat spacetime. Specifically, we found that the thermal correction of the renormalized Casimir free energy gradually decreases as the temperature increases, as shown in figure \ref{figfree}. We have also studied the renormalized thermal Casimir free energy as a function of the plate's distance in SI units (figure \ref{FrentoL}). With the parameter choice values, the allowed temperature can be constrained in the range $0 \le T\le 10^5$ K. We found that as the temperature increases, the magnitude of the renormalized Casimir free energy increases and vice versa.

Moreover, the thermal correction of the others  thermodynamics quantities can be easily obtained from the renormalized Casimir free energy. We found that the thermal corrections of the renormalized Casimir entropy and internal energy exhibit a similar behavior, where the thermal corrections increase monotonically as the temperature increases, then approach finite asymptotic values. In particular, at beginning of the plots (figure \ref{fig3}), the Casimir entropy has a higher rise compared to that of the internal energy, due to an extra one-power of $\tilde{\beta}$ in the expression of $\mathcal{U}^{\text{ren}}$, comparing both Eqs.\eqref{sren} and \eqref{uren}. As for the thermal correction of the renormalized Casimir heat capacity, it increases monotonically as the temperature becomes large, reaches a maximum, then gradually decreases and approaches zero. In the low-temperature limit, these thermal corrections contain several contributions: a power-law term proportional to ($\tilde{\beta}^{- 3}$) from modified low-frequency Matsubara modes, a term proportional to ($\tilde{\beta}^{- 4}$) from blackbody radiation confined within the cavity volume, and an exponentially suppressed term ($e^{- 2 \pi m \tilde{\beta}}$). Lastly, as the temperature approaches zero, the Casimir internal energy $\mathcal{U}^{\tmop{ren}}$ goes to $E_{\text{Cas,0}}$ and the Casimir entropy $\mathcal{S}^{\tmop{ren}}$ is zero, which is consistent with the third law of thermodynamics.

For future work, several extensions of this study may be of interest. These include considering different boundary conditions, massive fields, or non-circular motion \cite{Dutta:2024luw,Ishkaeva:2026tot}, as well as considering the Casimir apparatus in magnetic field background \cite{Ostrowski:2005rm,Elizalde:2002kb,Cougo-Pinto:1998jun,Erdas:2013jga,Rohim:2023tmy,Erdas:2020ilo}, exploring configurations beyond the zero–tidal-force wormhole model \cite{Kuhfittig:2014}, for rotating wormhole \cite{Teo:1998, Garattini} and Neutron star wormhole \cite{deFarias:2025vks} with a similar setup. It would be worthwhile to investigate whether global geometric effects \cite{Zhang:2017} or strong-field regimes can lead to observable deviations in finite-temperature Casimir phenomena \cite{Giacomelli:2024}.

\subsection*{Acknowledgments}
A. Romadani thanks the National Research and Innovation Agency (BRIN) for the funding support of the Degree by Research (DbR) at Sepuluh Nopember Institute of Technology (ITS). The work of A.S.A is supported by an appointment to the Young Scientist Training (YST) Program at the APCTP through the Science and Technology Promotion Fund and Lottery Fund of the Korean Government and is also supported by the Korean Local Governments - Gyeongsangbuk-do Province and Pohang City.

\subsection*{Conflict of interest declaration}
The authors declare no conflicts of interest regarding this study.

%\appendix
%%%%%%%%%%%%%%%%%%%%%%%%%%%%%%%%%%%%%%%%%%%%
%Appendix A
%%%%%%%%%%%%%%%%%%%%%%%%%%%%%%%%%%%%%%%%%%%%
%\section{Appendix A}
%\bibliographystyle{utphys}
%\bibliography{bibliography}

\end{document}